\documentclass[aps,prl,final,twocolumn,showpacs]{revtex4}

\usepackage{amsmath,amssymb,graphicx}

\begin{document}

\title{Recent progress in mathematical diffraction}

\author{Uwe Grimm$^{a}$ and Michael Baake$^{b}$}
\affiliation{
{}$^{a}\!$Department of Mathematics and Statistics,
The Open University,
Walton Hall, Milton Keynes MK7 6AA, UK\\
{}$^{b}\!$Fakult\"{a}t f\"{u}r Mathematik, 
Universit\"{a}t Bielefeld,
Postfach 100131, 33501 Bielefeld, Germany}

\begin{abstract}
  A brief summary of recent developments in mathematical diffraction
  theory is given. Particular emphasis is placed on systems with
  aperiodic order and continuous spectral components. We restrict
  ourselves to some key results and refer to the literature for
  further details.
\end{abstract}


\maketitle

\section{Introduction}

Diffraction methods \cite{C} continue to provide the main tool for the
structure analysis of solids. The corresponding inverse problem of
determining a structure from its diffraction is difficult and, in
general, does not define a structure uniquely.  Kinematic diffraction,
an approximation that is reasonable for X-ray diffraction where
multiple scattering effects can be neglected, is well suited for a
mathematical approach via \emph{measures}. Measures are
generalisations of the classic concept of Lebesgue measure used in
volume integration and provide a natural mathematical concept to
quantify the distribution of matter in space as well as the
distribution of scattering intensity. This mathematical approach to
diffraction was pioneered by Hof \cite{Hof} and has substantially been
developed since the discovery of quasicrystals required an extension
of the methods used to compute the diffraction of perfectly periodic
crystals.

The need for further insight emerged from the question of which
distributions of matter, beyond perfectly periodic crystals, lead to
pure point diffraction patterns, hence to diffraction patterns
comprising sharp Bragg peaks only. More recently, it has become
apparent that one also has to study continuous diffraction in more
detail, with a careful analysis of the different types (singular and
absolutely continuous) of diffuse scattering involved. This is both of
interest from a mathematical point of view, since the diffraction
spectrum is closely related to the dynamical spectrum of the
associated dynamical system \cite{dyn}, as from an experimental point
of view, where diffuse scattering including candidates for singular
continuous scattering are observed \cite{Withers}.

In this brief account of a tutorial review, we summarise key results,
putting particular emphasis on the analysis of non-periodic
structures. Following the presentation in our recent review articles
\cite{BG11,BG12}, general results are introduced and discussed on the
basis of various characteristic examples, with minimal use of formal
arguments or proofs. For details and more background material, we
refer to the comprehensive treatment in \cite{tao}.

\section{Diffraction measure}

We consider the diffraction measure mainly for Delone sets
$\varLambda\subset\mathbb{R}^{d}$, which are point sets where the
points neither get arbitrarily close nor so sparse that they
accommodate arbitrarily large empty balls. Define the corresponding
weighted \emph{Dirac comb}
\begin{equation}\label{eq:comb}
    \omega \, = \, 
    w\, \delta_{\varLambda}\, = \, 
    \sum_{x\in\varLambda} w(x)\, \delta_{x}\, ,
\end{equation}
where $\delta_{\varLambda}=\sum_{x\in\varLambda}\delta_{x}$ denotes
the uniform comb of point scatterers $\delta_{x}$ at all positions
$x\in\varLambda$, and where $w(x)\in\mathbb{C}$ represents the
scattering weight at position $x$. In most cases we are interested in
(and all that appear below), the function $w$ is such that $\omega$ is
a translation bounded measure on $\mathbb{R}^{d}$. For instance, if
$\varLambda$ is Delone, it suffices if $w$ is a bounded function.

The natural \emph{autocorrelation measure} of $\omega$ is defined as
the limit (provided it exists)
\begin{equation}\label{eq:def-gamma}
   \gamma \, =  \, \gamma^{}_{\omega} 
                  = \, \omega \circledast \widetilde{\omega} 
                 \, := \lim_{R\to\infty}
                \frac{\;\omega|^{}_{R} \ast \widetilde{\omega|^{}_{R}}\;}
                   {\mathrm{vol} (B_R)} ,
\end{equation}
where $\omega|^{}_{R}$ denotes the restriction of $\omega$ to the open
ball $B_{R}$ of radius $R$ around $0\in \mathbb{R}^{d}$, and
$\widetilde{\mu}$ is the `flipped-over' version of a measure $\mu$
defined by $\widetilde{\mu}(g) = \overline{\mu(\widetilde{g})}$ for
any measurable function $g$, where
$\widetilde{g}(x)=\overline{g(-x)}$. The operation $\circledast$ (also
known as the \emph{Eberlein convolution}) is a volume-averaged
analogue of the ordinary convolution $\ast$ of measures.

Assuming that the autocorrelation measure $\gamma$ of $\omega$ exists,
its Fourier transform $\widehat{\gamma}$ is a well-defined translation
bounded, positive measure, called the \emph{diffraction measure} of
$\omega$.  It describes the kinematic scattering intensity observed in
an experiment. The measure $\widehat{\gamma}$ has a unique decomposition
\begin{equation}\label{eq:decomp}
   \widehat{\gamma} \; = \; 
            \widehat{\gamma}^{}_{\mathrm{pp} } +
            \widehat{\gamma}^{}_{\mathrm{sc} } +
            \widehat{\gamma}^{}_{\mathrm{ac} }
\end{equation}
into a pure point part (comprising the Bragg peaks, of which there are
at most countably many), an (with respect to Lebesgue measure)
absolutely continuous part (the diffuse background scattering, which
has a locally integrable density) and a singular continuous part
(which comprises anything that remains).

\section{Perfect crystals}

A perfect (infinite) crystal in $\mathbb{R}^{d}$ corresponds to a
lattice-periodic discrete structure. Denoting its lattice of periods
by $\varGamma\subset\mathbb{R}^{d}$ and specifying the decoration of a
fundamental domain of $\varGamma$ with scatterers by a finite measure
$\mu$, one obtains the crystallographic measure
\begin{equation}\label{eq:cryst}
    \omega\, =\, \mu \ast \delta^{}_{\varGamma}\, .
\end{equation}
with autocorrelation $\gamma = \mathrm{dens} (\varGamma)\ (\mu \ast
\widetilde{\mu}) \ast \delta^{}_{\varGamma}$.
The Fourier transform of lattice-periodic measures can be calculated using 
Poisson's summation formula
\begin{equation}\label{eq:psf}
 \widehat{\delta^{}_{\varGamma}}
     \, = \, \mathrm{dens} (\varGamma) \, 
     \delta^{}_{\varGamma^{*}}\, ,
\end{equation}
where $\varGamma^{*}$ denotes the \emph{dual} or
\emph{reciprocal lattice} of $\varGamma$,  defined by
\[
    \varGamma^{*} \, =\, \{ x\in\mathbb{R}^{d} \mid \mbox{$\langle x|
      y\rangle \in \mathbb{Z}$ for all $y \in \varGamma$} \} \, .
\]
Note that our convention for the Fourier transform is $\widehat{\phi}
(k) \, := \int_{\mathbb{R}^d} e^{-2\pi i \langle k| x\rangle} \, \phi
(x)\, \mathrm{d}x$, where $k,x\in\mathbb{R}^{d}$ with scalar product
$\langle k|x\rangle$. The diffraction measure of the crystallographic
measure of Eq.~\eqref{eq:cryst} is then obtained as
\begin{equation}\label{eq:crystdiff}
     \widehat{\gamma} \,= \, \bigl(  \mathrm{dens} (\varGamma) \bigr)^{2}
     \, \big| \widehat{\mu} \big|^{2} \, \delta^{}_{\varGamma^{*}} \, ,
\end{equation}
which is a pure point measure supported on the reciprocal lattice
$\varGamma^{*}$, with scattering intensities that can be calculated
from the Fourier transform of the finite measure $\mu$.

As a simple example, consider a $\mathbb{Z}^{2}$-periodic system with
two scatterers in a unit cell, one of unit scattering strength at
position $(0,0)$ and one of scattering strength
$\alpha\in\mathbb{C}$ at position $(a,b)$.  The corresponding weighted
Dirac comb is $\omega = \mu * \delta^{}_{\mathbb{Z}^{2}}$ with $\mu =
\delta_{(0,0)} +\alpha \delta_{(a,b)}$. The diffraction measure is 
$\widehat{\gamma^{}_{\omega}}=\lvert\widehat{\varrho}\,\rvert^{2}
\delta^{}_{\mathbb{Z}^{2}}$ (note that the lattice $\mathbb{Z}^{2}$ is
self-dual) with diffraction intensities
\[
   I(k_{1},k_{2}) \, = \, \lvert\widehat{\varrho}\,\rvert^{2}(k_{1},k_{2}) 
   \, = \, \bigl| 1 + \alpha e^{- 2 \pi i (k_{1} a + k_{2} b)} \bigr|^{2}\, ,
\]
which are evaluated at the points $(k_{1},k_{2})\in\mathbb{Z}^{2}$ of
the reciprocal lattice. Note that, in general, the intensity
distribution is not periodic, while the location set of the Bragg 
peaks is (provided there are no extinctions).

\section{Euclidean model sets}

Model sets (or cut and project sets) arise from projections from a
lattice $\mathcal{L}$ in a higher-dimensional space. The general
setting for a Euclidean model set is encoded in the \emph{cut and
  project scheme} (CPS)
\begin{equation}\label{eq:cps}
\renewcommand{\arraystretch}{1.2}\begin{array}{r@{}ccccc@{}l}
   & \mathbb{R}^{d} & \xleftarrow{\,\;\;\pi\;\;\,} 
         & \mathbb{R}^{d} \times \, \mathbb{R}^{m}\!  & 
        \xrightarrow{\;\pi^{}_{\mathrm{int}\;}} & \mathbb{R}^{m} & \\
   & \cup & & \cup & & \cup & \hspace*{-2ex} 
   \raisebox{1pt}{\text{\footnotesize dense}} \\
   & \pi(\mathcal{L}) & \xleftarrow{\; 1-1 \;} & \mathcal{L} & 
   \xrightarrow{\; \hphantom{1-1} \;} & 
       \pi^{}_{\mathrm{int}}(\mathcal{L}) & \\
   & \| & & & & \| & \\
   &  L & \multicolumn{3}{c}{\xrightarrow{\qquad\quad\quad \;\;
       \;\star\; \;\; \quad\quad\qquad}} 
       &   {L_{}}^{\star} & \\
\end{array}\renewcommand{\arraystretch}{1}
\end{equation}
where $\mathbb{R}^{d}$ is the physical and $\mathbb{R}^{m}$ the
internal space, and $\mathcal{L}\subset \mathbb{R}^{d+m}$. The
associated natural projections are denoted by $\pi$ and
$\pi^{}_{\mathrm{int}}$. The bijectivity of the projection on
$L=\pi(\mathcal{L})\subset\mathbb{R}^{d}$ and the denseness of
$L^{\star}=\pi^{}_{\mathrm{int}}(\mathcal{L})\subset\mathbb{R}^{m}$
ensure that the $\star\,$-map $x\mapsto x^{\star}$ is well-defined on
$L$.

For a fixed CPS and a \emph{window} $W\subset\mathbb{R}^{m}$, the set
\begin{equation}\label{eq:ms}
    \varLambda \, = \,
    \bigl\{  x\in L \mid  x^{\star} \in W \bigr\} ,
\end{equation}
is called a \emph{model set} (or cut and project set). As long as the
window is sufficiently well behaved, see \cite[Sec.~9.4]{tao} for the
details, the corresponding Dirac comb $\delta_{\varLambda}$ has the
pure point diffraction measure
\begin{equation}\label{eq:modeldiff}
    \widehat{\gamma}\,  = \sum_{k\in L{}_{}^{\circledast}}
         \lvert A(k) \rvert^{2}\, \delta_{k}\, ,
\end{equation}
which is supported on the Fourier module $L^{\circledast} = \pi
(\mathcal{L}^{*})$, the projection of the higher-dimensional dual
lattice.  The diffraction amplitudes $A(k)$ are explicitly given by
\begin{equation}\label{eq:modelamp}
   A(k) \, = \, 
  \frac{\mathrm{dens} (\varLambda)}{\mathrm{vol} (W)}
  \, \widehat{1^{}_{\! W}} (-k^{\star})\, ,
\end{equation}
where $1^{}_{W}$ denotes the characteristic function of the window
$W$, and $\star$ is the star-map of the CPS. So, the calculation of
diffraction intensities essentially requires the Fourier transform of
the characteristic function of the window $W$, which can be done
explicitly for many examples with polygonal or spherical
windows. Clearly, the symmetry of the window $W$ manifests itself
in the symmetry of the diffraction intensities; see \cite{tao}
for examples.

Note that, while $\widehat{\gamma}$ is a pure point measure, it is
supported on the projection of the entire reciprocal lattice, which
(in general) results in a dense point set in
$\mathbb{R}^{d}$. Nevertheless, the total intensity scattered into any
region of space, which corresponds to summing up infinitely many
intensities of peaks in that region, always remains
finite. Restricting to peaks with intensities above any given
threshold thus produces a discrete pattern of peaks.

\begin{figure}
\centerline{\includegraphics[width=\columnwidth]{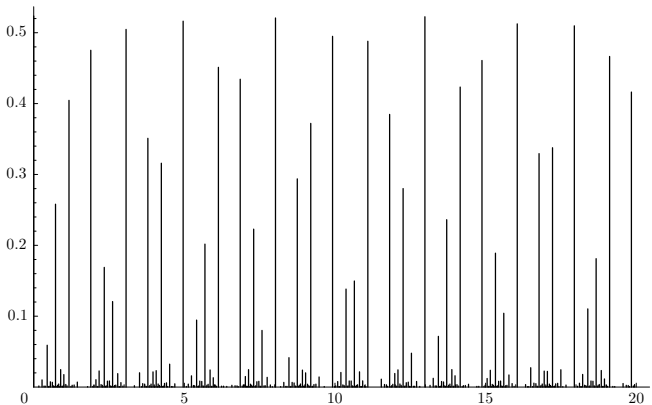}}
\caption{Sketch of the diffraction  of the Fibonacci point set for $0\le k\le
20$. A peak of this pure point measure is represented as a line with a
height that equals its intensity.  All peaks with at least $1/1000$ of
the central intensity $I(0)=(\tau+1)/5\approx 0.5236$ are included.}
\label{fig:fibodiff}
\end{figure}

As a one-dimensional example, consider the Fibonacci point set which
is obtained from a CPS with planar lattice $\mathcal{L} =
\bigl\{(x,x^{\star})\mid
x\in\mathbb{Z}[\tau]\bigr\}\subset\mathbb{R}^{2}$. Here,
$\tau=(1+\sqrt{5})/2$ is the golden number,
$L=\mathbb{Z}[\tau]=\{a+b\tau\mid a,b\in\mathbb{Z}\}$ and the star-map
acts as algebraic conjugation $(a+b\tau)^{\star}=a+b(1-\tau)$. The
window is $W=(-1,\tau-1]$ (chosen as half-open to avoid singular
cases), which produces a regular model set of density
$\tau/\sqrt{5}=(\tau+2)/5$.  The Fourier module is
$L_{}^{\circledast}=L/\sqrt{5}$, and the diffraction intensity for
$k\in L/\sqrt{5}$ is obtained as
\begin{equation}\label{eq:Fibo-int}
   I(k) \,=\, 
  \biggl(\frac{\tau}{\sqrt{5}}\,
  \mathrm{sinc} \bigl(\pi \tau {k_{}}^{\star}\bigr)\biggr)^{2} ,
\end{equation}
where $\mathrm{sinc}(x)=\sin(x)/x$; a sketch of the diffraction
patterns in shown in Figure~\ref{fig:fibodiff}. The intensity function
$I(k)$ vanishes on $L_{}^{\circledast}$ if and only if $\tau
{k_{}}^{\star} \in \mathbb{Z} \setminus \{ 0 \}$. This corresponds to
$k = \ell \tau$ with $\ell \in \mathbb{Z} \setminus \{ 0 \}$, and the
Bragg peaks at these positions are extinct. Another apparent feature
of Figure~\ref{fig:fibodiff} is the presence of series of peaks with
increasing intensity. These series appear at $\tau$-scaled positions,
and are a consequence of the fact that
$\lvert{(\tau^m)}^{\star}\rvert=\lvert 1-\tau\rvert^{m}<1$, which by
Eq.~\eqref{eq:Fibo-int} implies that, for any $k$ with $I(k)>0$, one
has $I(\tau^{m}k)\rightarrow I(0)$ for $m\to\infty$.  In particular,
one can clearly see this phenomenon for the peaks at $\tau^i/\sqrt{5}$
with $0\le i\le 7$, where $\tau^5/\sqrt{5}\approx 4.96$.

\section{Singular continuous diffraction}

Singular continuous measures are rather strange, as they give no
weight to any single point, but are still concentrated to an
uncountable set of zero Lebesgue measure. A well-known example is the
'Devil's staircase', which is the distribution function of the
probability measure for the classic middle-thirds Cantor set, which is
constant almost everywhere. Singular continuous diffraction can occur
in experiment \cite{Withers} and in realistic models, and thus should
not be disregarded.

\begin{figure}
\centerline{\includegraphics[width=\columnwidth]{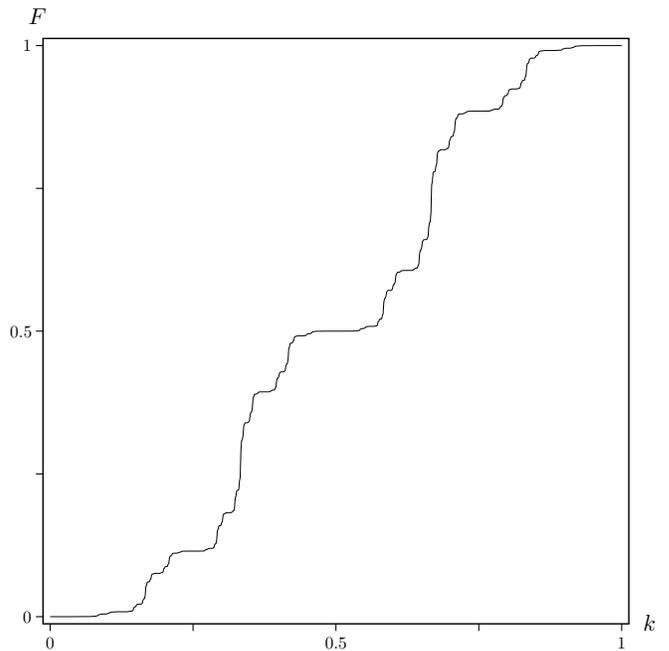}}
\caption{The (strictly increasing) distribution function $F(k)$ of the
  TM diffraction measure of Eq.~\eqref{eq:TM}.}
\label{fig:tmmeas}
\end{figure}

The paradigm for singular continuous diffraction is the Thue--Morse
(TM) system \cite{Kaku,Q,ME}, based on the binary substitution rule
$1\mapsto 1\bar{1}$, $\bar{1}\mapsto\bar{1}1$. The recursion
$v^{(n+1)}= v^{(n)} \bar{v}^{(n)}$ with initial condition $v^{(0)}=1$
clearly converges to the one-sided fixed point $v= v^{}_{0} v^{}_{1}
v^{}_{2} \cdots$ of the TM substitution.  The exponential sum
\begin{equation}\label{eq:sum}
    g_{n}(k) \, = \, \sum_{\ell=0}^{2^{n}-1} v^{}_{\ell}\, e^{-2\pi ik \ell},
\end{equation}
is the Fourier transform of the weighted Dirac comb
$\omega_{n}=\sum_{\ell=0}^{2^{n}-1}v_{\ell}\delta_{\ell}$ for the
(finite) word $v_{n}$, where we identify $\bar{1}=-1$. The exponential
sum of Eq.~\eqref{eq:sum} satisfies the recursion
\[
   g_{n+1}(k) \, = \, \bigl(1- e^{-2\pi ik 2^{n}}\bigr)\, g_{n}(k)
\]
for $n\ge 0$, with $g^{}_{0}(k)=1$. With $\widehat{\gamma} =
\lim_{n\to\infty} \lvert g_{n} \rvert^{2} / 2^{n}$, which converges
(as a measure) in the vague topology, the diffraction measure of the
TM system can now be represented as a \emph{Riesz product} \cite{Q}
\begin{equation}\label{eq:TM}
   \widehat{\gamma}\, = 
     \prod_{n\ge 0} \bigl( 1 - \cos(2^{n+1}\pi k)\bigr).
\end{equation}
The corresponding distribution function $F(k):=
\widehat{\gamma}([0,k])$ is continuous. Moreover, it possesses a
uniformly converging Fourier series \cite{BG08,BGG}; it is shown in
Figure~\ref{fig:tmmeas}.  For scaling properties of the Thue--Morse
spectrum, we refer to \cite{BGN} and references therein.

While singular continuous spectra may appear rather special, they are
in fact quite common; compare \cite{Knill} for a corresponding
concrete result from the spectral theory of Schr\"{o}dinger
operators. The TM approach can be generalised to show that large
classes of bijective binary substitution systems have purely singular
continuous diffraction \cite{Nat}, which includes higher-dimensional
systems such as the squiral tiling \cite{squiral}.

\section{Absolutely continuous diffraction}

Usually, absolutely continuous diffraction is connected with disorder
\cite{BLR}. Indeed, randomness often gives rise to continuous
components in the diffraction, though it is important to remember that
there are also `ordered' structures that may lead to absolutely
continuous spectra. The probably best known example is the
Rudin--Shapiro (RS) system, which can be defined by the sequence of
weights $w_{n}\in\{\pm 1\}$ with initial conditions $w(-1)=-1$,
$w(0)=1$, and the recursion
\begin{equation}\label{eq:rs}
   w_{4n+\ell}=
    \begin{cases} w_{n},  & \mbox{for $\,\ell\in\{0,1\}$,} \\
          (-1)^{n+\ell}\,w_{n}, & \mbox{for $\,\ell\in\{2,3\}$.}
     \end{cases}
\end{equation}
The corresponding Dirac comb $\omega=\sum_{n\in\mathbb{Z}}
w_{n}\delta^{}_{n}$ has the autocorrelation measure $\gamma =
\delta^{}_{0}$, and hence Lebesgue measure as diffraction measure. In
other words, the diffraction is completely `featureless', with constant
intensity for all $k\in\mathbb{R}$.

A simple random system that shares this property is the binary
Bernoulli chain, where weights $v_{n}$ for $n\in\mathbb{Z}$ are chosen
independently to be either $1$ or $-1$, with equal probability
$p=1/2$. These two systems are thus homometric. More surprisingly, as
shown in \cite{BG09}, this example can be generalised to an entire
family of homometric systems of the form
\[
 \omega_{p} \, = \sum_{n\in\mathbb{Z}} w_n\, X_n\,\delta_{n}\, .
\]
where $(w_{n})^{}_{n\in\mathbb{Z}}$ is the binary RS system defined in
Eq.~\eqref{eq:rs}, and where $(X_n)^{}_{n\in\mathbb{Z}}$ is an i.i.d.\
family of random numbers taking values $1$ and $-1$ with probabilities
$p$ and $1-p$. This example should serve as a warning concerning the
inverse problem of diffraction --- in general, this is not unique, and
may not even be sensitive to `order' or `disorder', for instance in the
sense of entropy (as in this case).

\begin{figure}
\centerline{\includegraphics[width=\columnwidth]{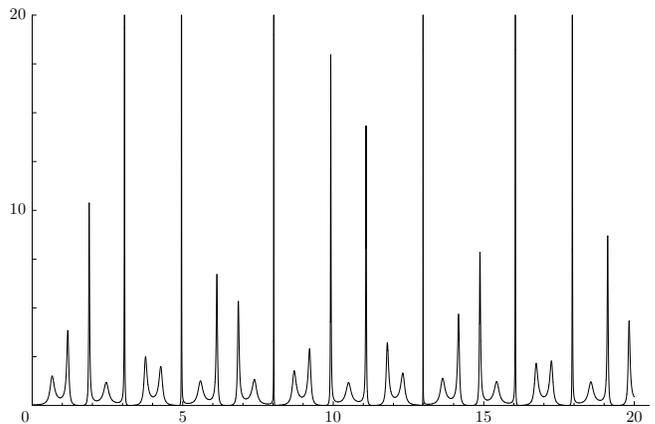}}
\caption{The absolutely continuous part of the diffraction pattern of
  a Fibonacci random tiling, for the same range of the wave number $k$
  as in Figure~\ref{fig:fibodiff}.}
  \label{fig:fibort}
\end{figure}

While some progress has been made to understand diffraction of systems
with stochastic disorder (see \cite{BG11,BG12,tao} and references
therein), systems with correlated disorder as well as random tilings,
which are of particular interest in the context of quasicrystals, are
generally difficult to treat. As a simple example, consider the
one-dimensional Fibonacci random tiling, which is the ensemble of all
tilings of $\mathbb{R}$ with two prototiles (one interval of length
$\tau$ and one of length $1$), chosen with probabilities $p=\tau^{-1}$
and $1-p$, respectively. For this system, you find that, almost
surely, the diffraction measure satisfies
 \begin{equation}\label{eq:fibort-diff}
    \widehat{\gamma} \; = \, 
     \biggl(\frac{\tau + 2}{5}\biggr)^{2} \delta^{}_{0}
    \, +\,  h(k) \, \lambda\, ,
\end{equation}
where $\lambda$ denotes Lebesgue measure and where the corresponding
Radon--Nikodym density $h$ is given by
\[
  h(k)  =  \frac{\tau+2}{5}\frac{(\sin (\frac{\pi k}{\tau}))^{2} }
      {\tau^{2} \, (\sin(\pi k \tau))^{2} + \tau \, 
            (\sin (\pi k))^{2} - (\sin (\frac{\pi k}{\tau}))^{2} }\, .
\]
Beside the trivial Bragg peak at $k=0$, the diffraction is absolutely
continuous. The function $h$ if shown in Figure~\ref{fig:fibort};
while it is smooth, it still displays a spiky structure resembling the
pure point diffraction of the perfect system shown in
Figure~\ref{fig:fibodiff}.

An alternative way of introducing disorder into a Fibonacci system (or
more generally, into a substitution system) is obtained by mixing
different inflation rules \emph{locally} \cite{GL,BM}. This produces
interesting tiling ensembles of positive entropy, where under certain
conditions (as in the case of the noble means substitutions) all
realisations are Meyer sets \cite{BM}.  Consequently, they possess
non-trivial Bragg diffraction (in line with general results by
Strungaru \cite{Nicu}), but in addition show absolutely continuous
diffraction; see \cite{BM,BMnew} for details.

\section{Outlook}

The discovery of quasicrystals in 1982 was based on its unusual
diffraction pattern, displaying crystallographically `forbidden'
icosahedral symmetry. Such diffraction patterns are by now well
understood, in the sense that it has been shown rigorously that, with
some modest assumptions on the window, the diffraction measure of
mathematical quasicrystals (model sets) are pure point measures, so
comprise Bragg peaks only. It remains to achieve a better
understanding of the cases with continuous diffraction, which carries
important information on the structure of a system as well, not just
in the case where Bragg peaks are absent. While progress has been
made, and some explicit examples of singular and absolutely continuous
spectra have become accessible, see \cite{BBM} and references therein,
there are still important systems that require further research. This
includes proper random tilings in two and more dimensions, as well as
substitution and inflation-based structures such as non-Pisot
substitution systems or pinwheel-type tilings with continuous
symmetries.

From the mathematical point of view, the relation between diffraction
spectra and dynamical spectra of the associated dynamical system
(under translation action) is now much better understood; see
\cite{dyn} and references therein for recent developments.
Also, with methods from the theory of (stochastic) point processes,
the inverse problem of structure determination from a pure
point (or Bragg) diffraction spectrum has been understood in
an abstract setting in rather large generality \cite{LM}; see also
\cite{TB,Terauds} for additional examples in this context. The
corresponding problem for general (mixed) spectrum is still open,
and appears to be rather challenging.

\section*{Acknowledgements}

This work was supported by the German Research Foundation (DFG) within
the CRC~701.

\end{document}